# Developing a High Performance Software Library with MPI and CUDA for Matrix Computations

**Bogdan Oancea[1][*], Tudorel Andrei[2]**

[1]"Nicolae Titulescu" University of Bucharest, e-mail: bogdanoancea@univnt.ro, Calea Văcărești, nr. 185, sector 4, București, Romania
[2]The Bucharest Academy of Economic Studies, e-mail: andreitudorel@yahoo.com, București, Romania

**Abstract**

*Nowadays, the paradigm of parallel computing is changing. CUDA is now a popular programming model for general purpose computations on GPUs and a great number of applications were ported to CUDA obtaining speedups of orders of magnitude comparing to optimized CPU implementations. Hybrid approaches that combine the message passing model with the shared memory model for parallel computing are a solution for very large applications. We considered a heterogeneous cluster that combines the CPU and GPU computations using MPI and CUDA for developing a high performance linear algebra library. Our library deals with large linear systems solvers because they are a common problem in the fields of science and engineering. Direct methods for computing the solution of such systems can be very expensive due to high memory requirements and computational cost. An efficient alternative are iterative methods which computes only an approximation of the solution. In this paper we present an implementation of a library that uses a hybrid model of computation using MPI and CUDA implementing both direct and iterative linear systems solvers. Our library implements LU and Cholesky factorization based solvers and some of the non-stationary iterative methods using the MPI/CUDA combination. We compared the performance of our MPI/CUDA implementation with classic programs written to be run on a single CPU.*

## 1. Introduction

From physics and engineering to macroeconometric modeling, solving large linear systems of equations is a common problem. Such problems rely on high performance computing. One of the parallel programming paradigms is the message passing with its implementation using the MPI library [Snir et al., 1996]. About ten years ago MPI clusters were the first choice for many scientific applications but nowadays GPUs are used for performing general computations. In 2003 [Harris et al., 2003] pointed out a new approach to obtain a high megaflop rate to the applications when he started to use GPUs (graphical processing unit) for non-graphics applications. Current Graphics Processing Units contain high performance many-core processors capable of very high FLOP rates and data throughput being truly general-purpose parallel processors. Since the first idea of Mark Harris, many applications were ported to use the GPU for compute intensive parts and they obtain speedups of few orders of magnitude comparing to equivalent implementations written for normal CPUs.

At this moment, there are several models for GPU computing: CUDA (Compute Unified Device Architecture) developed by NVIDIA [NVIDIA, 2011], Stream developed by AMD [AMD, 2008] and a new emerging standard, OpenCL [Khronos, 2009] that tries to unify different GPU general computing API implementations providing a general framework for software development across heterogeneous platforms consisting of both CPUs and GPUs.

Combining the message passing based clusters with the very high FLOP rates of GPUs is a relatively recent idea [Fengshun et al., 2012]. We developed a hybrid linear algebra library that uses both MPI for spreading the computations among the computing nodes in a cluster and CUDA for performing the local computations on each node of the cluster. Thus, our library exploits a complex memory hierarchy: a distributed memory among the computing nodes in the cluster and a shared memory on each node which is, in fact, the device memory of the local GPUs.

---

[*] Corresponding author: bogdanoancea@univnt.ro

## 2. Serial iterative and direct methods

Stationary iterative methods such as Jacobi and Gauss-Seidel are well known and there are many textbooks that describe these methods [Golub and Van Loan, 1996]. An alternative to the stationary methods are Krylov techniques which use information that changes from iteration to iteration. Operations involved in Krylov methods are inner products, saxpy and matrix-vector products that has the complexity of $O(n^2)$, making them computational attractive for large systems of equations. One of the most used Krylov' method is the conjugate gradient (CG) [Golub and Van Loan, 1996] which solves SPD systems and in exact arithmetic gives the solution for at most $n$ iterations.

A relatively new method for general non symmetric linear systems is the Generalized Minimal Residuals (GMRES) introduced by [Saad, 1996]. GMRES uses a Gram-Schmidt orthogonalization process and requires the storage and computation of an increasing amount of information at each iteration. These difficulties can be alleviated by restarting the computations after a fixed number of iterations. The intermediate results are then used as a new initial point.

Another non-stationary method is the BiConjugate Gradient (BiCG). BiCG generates two mutually orthogonal sequences of residual vectors and A-orthogonal sequences of direction vectors. The updates for residuals and for the direction vectors are similar to those of the CG method, but are performed using system's matrix and its transpose. In our library we've implemented a version of BiCG called BiCGSTAB.

The alternative to the iterative methods for solving a linear system $Ax = b$ is the *direct method* that consists in two steps:

- The first step consists in matrix factorization: $A = LU$ where $L$ is a lower triangular matrix with 1s on the main diagonal and $U$ is an upper triangular matrix. In the case of SPD matrices, we have $A = LL^t$.
- In the second step we have to solve two linear systems with triangular matrices: $Ly = b$ and $Ux = y$.

The standard LU factorization algorithm with partial pivoting is given in [Golub and Van Loan, 1996]. The computational complexity of this algorithm is $\Theta(2n^3/2)$. After computation of the matrix factors L and U we have to solve two triangular systems: $Ly = b$ and $Ux = y$ These systems are solved using forward and backward substitution with a computational complexity of $\Theta(n^2)$, the most important computational step being the matrix factorization.

Computers with memory hierarchies are used more efficiently if the matrix factorization uses BLAS Level 3 operations [Dongarra et al., 1990] besides level 1 and level 2 operations [Lawson et al., 1979], [Dongarra et al., 1988]. It is well-known, level 3 BLAS operations have a better efficiency than level 1 or level 2 operations. The standard way to change a level 2 BLAS operations into a level 3 BLAS operation is delayed updating. In the case of the LU factorization algorithm we will replace k rank 1 updates with a single rank k update resulting a block algorithm. A detailed description of the block LU factorization algorithm is given in [Oancea, 2003].

## 3. The implementation of parallel algorithms

The serial algorithms presented here may not always be appropriate for very large matrices, parallel versions being more suitable for such matrices.

Software packages for solving linear systems have known a powerful evolution. A software package for linear algebra problems that emerged as a de-facto standard was LAPACK [Anderson et al., 1992] which was adapted for parallel computation resulting ScaLAPACK [Choi et al., 1992] library. Many other software packages for parallel computation have been developed so far: PETSc [Balay et al., 2004] PARPACK [Maschhoff and Sorensen, 1996], SuperLU [Xiaoye and Demmel, 2003].

Since the introduction of GPU general computation frameworks (CUDA, and Stream) many numerical libraries were ported to them: CUBLAS [NVIDIA, 2007] is a CUDA implementation of the BLAS library, MAGMA [Horton and Dongarra, 2001] is a collection of next generation linear algebra GPU accelerated libraries for heterogeneous GPU-based architectures, CULA [CULA, 2012] is a library that provides an accelerated implementation of the LAPACK and BLAS libraries for both dense and sparse linear algebra.

Previously [Oancea and Zota, 2003] and [Oancea et al. 2011] we presented a library that implements parallel algorithms for linear systems solving - PLSS (Parallel Linear System Solver). The PLSS library was designed with an easy to use interface almost identical with the serial algorithms' interface. Now, we improved this library combining the distributed computing used in PLSS with CUDA accelerated local computations. We named the new library CUPLSS.

The library has a very simple interface that makes the software developing process very easy because the parallelism is hidden from the user. This goal was obtained by means encapsulation of data and distribution and communication in opaque objects that hide the complexity from the user. Our library was developed in C and we used MPICH implementation of the MPI for the communication between processors. The local computations on each MPI node is further accelerated using CUDA, so that each local call of a computational intensive kernel is sent to be executed on the GPU device. The simplified structure of the computing architecture used for our tests in presented in Figure 1.

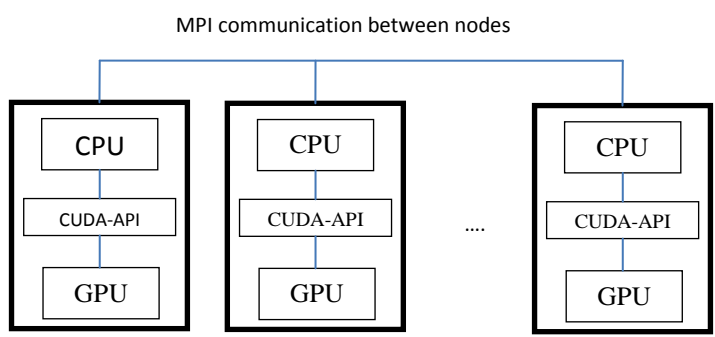

**Fig. 1**. MPI – CUDA hybrid architecture

MPI is used to facilitate the communication between nodes and exploit coarse grained parallelism of the applications and CUDA accelerates local computations on each node exploiting the fine grained parallelism. In our experiments we used a cluster of 16 workstations each having an Intel QuadCore Q6600 processor and NVIDIA GeForce GTX 280 GPU. The communication between nodes is achieved using a standard Gigabit LAN.

Our library is structured on four levels, as we can see in Figure 2.

| Application Program Interface – provides routines for parallel linear system solving | | API level |
|---|---|---|
| Object manipulation routines | | Data distribution and encapsulation level |
| Data distribution level | | |
| Interface CUPLSS-MPI | Interface CUPLSS-CUBLAS | Architecture independent level |
| Native MPI library | | Architecture dependent level |
| Native CUBLAS library | | |
| CUDA runtime | | |

**Fig. 2**. CUPLSS structure

The first level contains the CUDA runtime, CUBLAS, MPI and C libraries which all are architecture dependent. The second level provides the architecture independence, which implements the interface between the first level and the rest of the CUPLSS package. The next level implements the data distribution model concentrating the details regarding distribution of vectors and matrices on processors.

The top level of the CUPLSS library is, the application programming interface. CUPLSS API provides a number of routines that implements parallel BLAS operations and parallel linear system solving operations: direct methods based on *LU* and Cholesky factorization and nonstationary iterative methods GMRES, BiCG, BiCGSTAB. The CUPLSS library uses a logical bidimensional mesh of processors (computing nodes). Wherever we used CUDA accelerated local operations the general flow of the computations was [Oancea, 2012]:

- **Step 1** : Allocate memory for matrices and vectors in the host memory;
- **Step 2** : Initialize matrices and vectors in the host memory;
- **Step 3 :** Allocate memory for matrices and vectors in the device memory;
- **Step 4** :Copy matrices from host memory to device memory;
- **Step 5:** Define the device grid layout:
  - Number of blocks
  - Threads per block
- **Step 6** : Execute the kernel on the device;
- **Step 7 :** Copy back the results from device memory to host memory;
- **Step 8**: Memory clean up.

## 4. Performance tests

We've tested our library for both single precision and double precision floating point numbers. For our tests we used a cluster of workstations connected through a 1000Mb Ethernet local network, each station having 4GB of main memory. The CUPLSS package uses the MPICH implementation of the MPI library and, for the local BLAS operations, uses the CUBLAS library that provides a high FLOP rate. Each node in the cluster is a computer with Intel Core2 Quad Q6600 processor running at 2.4 Ghz, 4 GB of RAM and a NVIDIA GeForce GTX 280 graphics processing unit (GPU) with 240 cores running at 1296 MHz, 1GB of video memory and 141.7 GB/sec memory bandwidth. The operating system used was Windows Vista 64 bit.

We have tested the CUPLSS package for both iterative and direct methods, for 1, 2, 4, 8, and 16 computing nodes. The dimension of the matrix was maintained fixed: 60000 rows and columns. Figure 3 shows the speedup of the parallel algorithms for the case when iterative methods are used to solve the model and figure 4 shows the speedup in the case of direct methods. The speedup is computed comparing the parallel algorithm with a serial version the uses one CPU. Both speedups are computed for single precision floating point numbers.

We wanted to evaluate how much CUDA accelerated local computation contributes to the overall performance. To achieve this goal we replaced all the calls to CUBLAS or other CUDA computations for local computations with calls to a serial BLAS implementation – ATLAS [Whaley et al. 2001] and calculated again the speedups. As figures 3 and 4 show, CUDA accelerated local computations improves the overall performance but this increase in the speedup is not very high. The main reason for this is the GPU memory contention on GPU device and the communication overhead incurred by the MPI processes that acts as synchronizing points between CUDA calls. The main advantage of using MPI and CUDA hybrid model is that it allows solving very large systems which could not fit in the GPU memory of one computer. Although a pure CUDA implementation of linear systems solvers shows very high speedups, very large matrices do not fit in the GPU memory so that distributing the matrices and using MPI message passing model is an advantage that cannot be neglected.

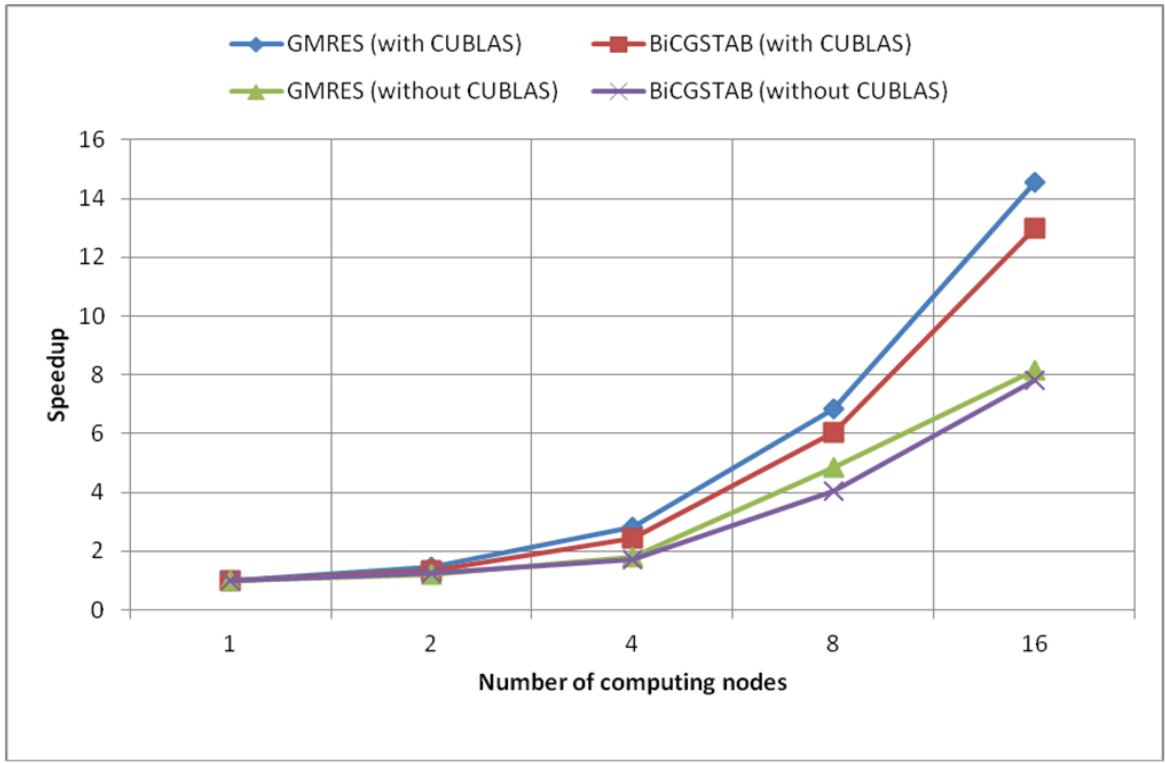

**Fig. 3**. The speedup for parallel versions of the iterative algorithms

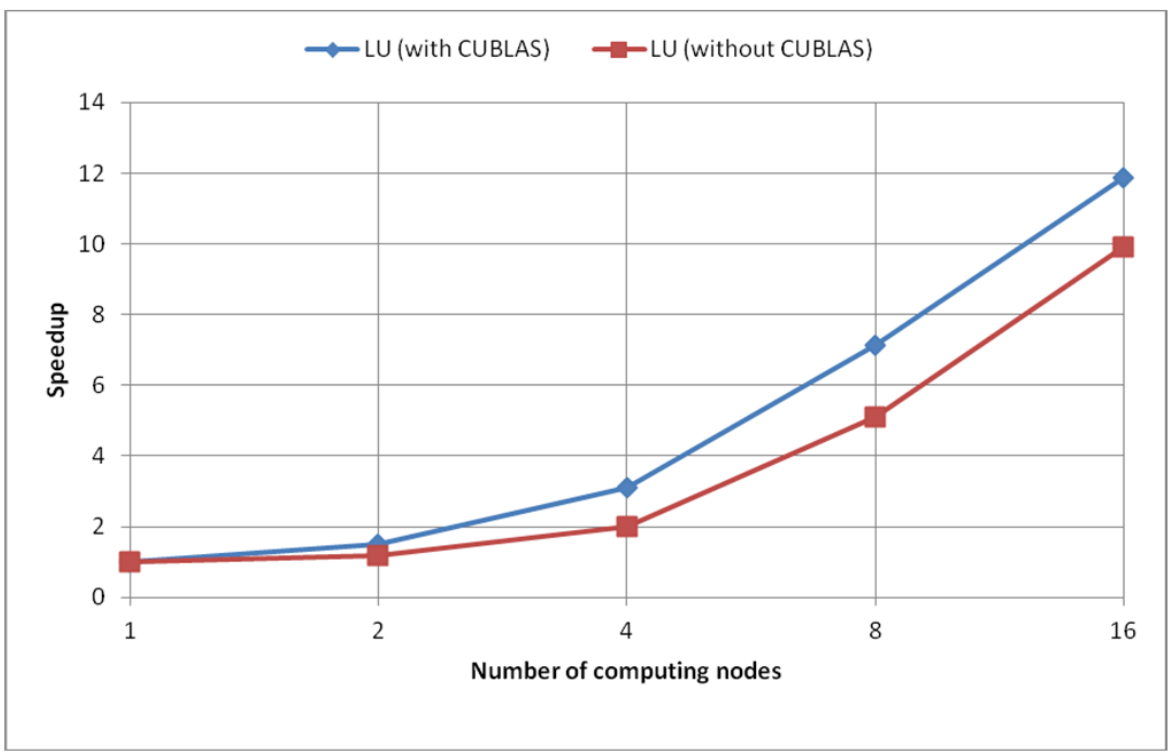

**Fig. 4**. The speedup for parallel versions of the LU factorization

## 5. Conclusions

We developed a hybrid MPI-CUDA library CUPLSS, that implements non-stationary iterative methods (GMRES, BiCGSTAB, BiCG) and direct methods for solving linear systems. We've made performance tests for our library in a network with 16 computing nodes and we obtained a good speedup. The speedup is higher for the methods based on matrix factorization compared with the iterative algorithms. We also tested how much CUDA accelerated local computation contributes to the overall performance by replacing all CUDA accelerated code with a serial code. The results shows that CUDA accelerated local computations improve the overall performance but the increase in performance is not very high mainly because of the GPU memory contention and MPI communication overhead.

In the future we intend to extend our library and to port it to OpenCL which will give hardware independence because CUDA is linked with NVIDIA devices.